# Is the Future of Materials Amorphous? Challenges and Opportunities in Simulations of Amorphous Materials

Ata Madanchi,[1] Emna Azek,[2] Karim Zongo,[3] Laurent K. Béland,[3] Normand Mousseau,[4] Lena Simine[2, *]

[1] Department of Physics, McGill University, Montréal, Québec, H3A 2T8, Canada

[2] Department of Chemistry, McGill University, Montréal, Québec, H3A 0B8, Canada.

[3] Department of Mechanical and Materials Engineering, Queen's University, Kingston, ON K7L 3N6, Canada

[4] Département de Physique, Institut Courtois and Regroupement Québécois sur les Matériaux de Pointe, Université de Montréal, Montréal, Québec H3C 3J7, Canada

**ABSTRACT:** Amorphous solids form an enormous and underutilized class of materials. In order to drive the discovery of new useful amorphous materials further we need to achieve a closer convergence between computational and experimental methods. In this review, we highlight some of the important gaps between computational simulations and experiments, discuss popular state-of-the-art computational techniques such as the Activation Relaxation Technique *nouveau* (ARTn) and Reverse Monte Carlo, and introduce more recent advances: machine learning interatomic potentials (MLIPs) and generative machine learning for simulations of amorphous matter. Examples are drawn from amorphous silicon and silica literature as well as from molecular glasses. Our outlook stresses the need for new computational methods to extend the time- and length- scales accessible through numerical simulations.

## Introduction

For over a century amorphous materials remained at the forefront of pure and applied research. From theoretical and computational perspective, at the forefront are the elusive physics of the glass transition[1], the quantification of hidden order in hyperuniform yet apparently random structures endowed with exotic properties[2], the interpretation of characterization experiments[3], the extension of computational simulations across multiple length- and time- scales for predictive modeling[4-7], and the development of machine learning approaches to help design useful amorphous materials. In this review, we present some of the challenges, the state-of-the-art computational approaches, and the opportunities that drive research farther into the amorphous chemical space.

Amorphous materials are typically defined as lacking long-range order, which means their atomic structure does not produce the well-defined diffraction patterns such as those seen in crystalline solids. This lack of a sharp, periodic arrangement results in indistinctive characteristics in many characterization experiments, such as broad diffuse scattering signals. However, this does not mean that amorphous structures are entirely without order. On a short-range scale, their atomic arrangements can display local order, with atoms adopting particular bonding configurations over a few atomic distances. This local order may vary depending on whether the material is in a liquid or solid state[8, 9]; the subset of amorphous structures in which the short-range order in the solid state is identical to that of the liquid phase is called glasses.

On the practical side, amorphous solids and glasses have numerous applications in biomedical engineering[10, 11], sports equipment[12], energy conversion[13, 14] and even nuclear waste immobilization through vitrification[15]. The number of potential glass compositions is estimated to be around $10^{52}$, with only about 200 compositions realized so far[16]. The search for new compounds in this vast chemical space is enabled by sophisticated computational methods ranging from computational modeling of materials and their properties to informing search policies and hypothesis generation.

The main challenge in modeling of amorphous materials with atomistic resolution is the disconnect between lab-based and computer-based molecular structures which primarily arises due to the limited length- and time-scales accessible to atomistic simulation methods. On the technical level this is due to (1) the difficult statistical sampling in the rugged energy landscapes characteristic of amorphous matter, and (2) the limited physical validity of interatomic interaction models. In Part 1 of this review, we outline some of the problems that arise because of the mismatch between simulations and experiments and list the key factors that lead to the mismatch. In Part 2, we review the evolution of domain-relevant sampling methodologies over the past few decades, and finally, in Part 3, we discuss the most recent advances that propel the field forward by integrating machine learning into the simulation protocols. We conclude in Part 4 with a brief partial summary of the remaining challenges that make the field of simulation of amorphous materials particularly exciting today.

**Part 1: Discord between computational simulations and experiments**

While massively parallel computing and, more recently, generative machine leaning models can, in principle, address the length-scale limitation of computational simulations, overcoming the timescale issue within the confines of molecular dynamics (MD) simulations— the workhorse of atomistic modeling—remains a significant challenge. Numerical methods can cover the first microsecond of reaction time, while experiments often take many orders of magnitude longer – seconds, minutes, hours. Consequently, various algorithms are employed to create computational models of amorphous materials, each exhibiting different structures and properties that often deviate significantly from those produced by experimental methods.

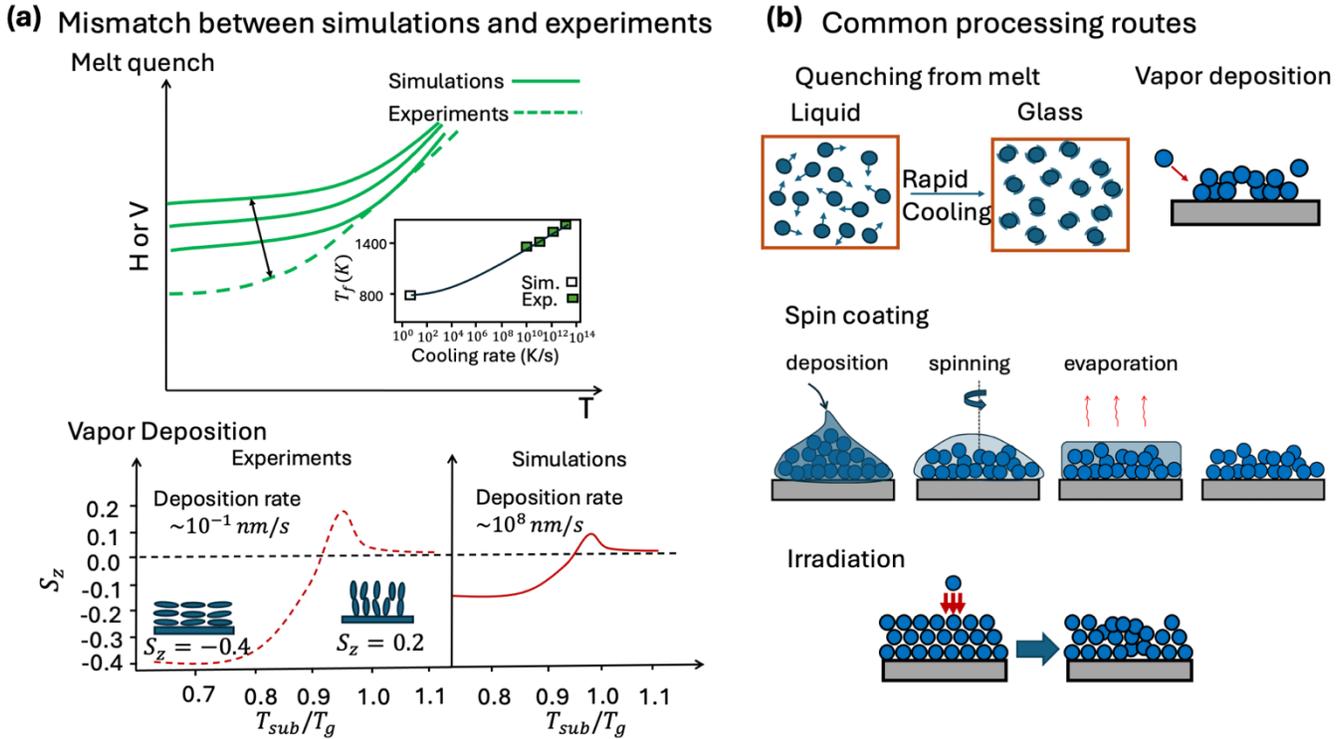

Figure 1. Summary of the existing simulation – experiment dichotomy in amorphous materials. (a) Mismatch between observable properties in simulated and real amorphous materials. (a) Top panel: Enthalpy H/Volume V as a function of temperature is schematically sketched for simulations (solid lines) and for experiments (dashed lines), illustrating isobaric cooling in a generic amorphous material as a function of temperature. Simulations that mimic experimental procedures typically employ cooling rates several order of magnitude faster ($10^{14}$–$10^{9}$ K/s) than those possible in traditional melt quench experiments ($10^2$–$10^0$ K/s)[17-20]. The inset shows fictive temperature, as a measure of stability, plotted versus cooling rate measured for silicate glasses in simulations and experiments (reproduced from[21]). Bottom panel: Similarly, simulated vapor deposition rates (often > $10^8$ nm/s)[22-27] outpace experimental rates which are usually below $10^0$ nm/s[28-34]. The figure shows the orientational order parameter, $S_z$, as a function of substrate temperature, observed in both simulations[35-37] and experiments[37, 38]. While simulations qualitatively capture the non-monotonic behavior observed in experiments there are qualitative discrepancies between simulated and experimental trends. (b) Illustrations of common processing routes for amorphous materials. Quenching from melt: A molten material is rapidly cooled (quenched) below its melting point, preventing crystallization and resulting in an amorphous solid; Vapor deposition: Atoms or molecules in a gaseous state are deposited onto a cold substrate. The rapid solidification on the surface prevents the formation of crystalline structures. Spin coating: A solution is deposited on a spinning platform where its self-assembly is governed by the interplay between surface tension, centripetal forces, and evaporation process. Irradiation: A crystalline/ceramic material is exposed to a flux of high-energy particles (e.g., ion, neutrons) resulting in an induced structural transformation.

The isobaric enthalpy (H) versus temperature graph in the top panel of Figure 1(a) illustrates the difference in the stability of the resulting structures obtained in melt-and-quench protocols placing the simulated samples in a qualitatively higher enthalpic register. The inset illustrates this point using another measure of stability – the fictive temperature which is plotted versus cooling rate measured for silicate glasses in simulations and experiments (reproduced from Ref. [21]) showing a large gap between the two.

The bottom panel in Figure 1(a) shows the behavior of order parameter $S_z$ that describes the average orientation of molecules within a thin film deposited via vapor deposition as a function of substrate temperature, observed in both simulations [35-37] and experiments[37, 38]. At high substrate temperatures ($T_{sub} > T_g$, where $T_g$ is the glass transition temperature), the molecules exhibit random orientations, resulting in $S_z \approx 0$. As the substrate temperature decreases below $T_g$ the molecules initially tend to align perpendicular to the substrate ($S_z > 0$). Upon further cooling, the molecular orientation transitions to a predominantly parallel alignment relative to the substrate ($S_z < 0$). While simulations qualitatively capture this non-monotonic behavior observed in experiments, certain limitations exist. Since, deposition rate has been found to be an effective parameter in dictating the relaxation dynamics of vapor deposited glasses[39], $T_g$ estimated from simulations is generally higher than experimentally determined values[40]. Such discrepancies arising from simpler models and limited time/length scales influence the quantitative agreement between simulated and experimental trends.

In Figure 1(b) we summarize schematically some of the typical processing routes starting from those that are reasonably approachable through simulations like melt-and-quench and vapor deposition to those that are notoriously difficult to model like spin-coating and irradiation. To illustrate the depth of the fundamental simulation-experiment discrepancy, consider for example that computational models of a-Si are routinely generated using melt-and-quench MD simulations, even though a-Si cannot be fabricated in the laboratory using a melt-and-quench process.

### Specific examples of systems of interest

Before we dive into the discussion of computational strategies, we first provide concrete context by introducing some of the commonly studied inorganic and organic amorphous solids, their applications, and the challenges associated with their preparation and simulation.

*Inorganic amorphous solids*

Amorphous silicon (a-Si) and amorphous silica (a-SiO$_2$) are canonical examples of non-metallic amorphous materials, with a-Si being a non-glassy amorphous solid and a-SiO$_2$ being a glassy one. Over more than 60 years of extensive research, a broad consensus has emerged regarding many of their characteristics. Both materials are considered continuous random networks with few coordination defects, existing as metastable phases possessing a free energy higher than that of their crystalline counterparts. However, their exact structure and properties are highly dependent on their processing history. From a technological perspective, this variability in structure and properties is significant for established industries, such as hydrogenated a-Si photovoltaic panels fabricated by chemical vapor deposition (CVD)[41], and critical for emerging industries, such as advanced photonics applications. Amorphous silicon, in particular, is being considered for use in photonic integrated circuits and advanced transistor devices because plasma-enhanced CVD a-Si deposition is compatible with other complementary metal–oxide–semiconductor fabrication steps, unlike traditional c-Si processing routes[42, 43]. For these applications, there is a direct relationship between processing conditions and electro-optical properties, making a-Si's suitability for advanced applications highly dependent on its fabrication method.

Similarly, a-SiO$_2$ exhibits variations when produced by different methods, such as fusing silica crystals versus formation through irradiation. Radiation-induced changes are particularly relevant for nuclear power plant aging and nuclear waste management. Irradiation can cause significant structural changes to silicate-based aggregates in concrete, leading to dimensional changes, alterations in chemical reactivity, and modifications of mechanical properties[44-48]. The variability in processing routes also raises fundamental questions about how to define a reference "perfect" form of a-Si and a-SiO$_2$. Different deposition conditions and thermal treatments can significantly affect the structure of a-Si[49-52]. Even a-Si created by ion-implantation, which is often considered as a good reference because of its low porosity, can be arbitrarily relaxed by thermal treatment[53, 54]. Likewise, there is clear evidence that irradiation of fused (vitreous) silica causes substantial structural changes[55-57].

*Molecular amorphous solids*

Beyond the traditional silicate-based compositions, when cooled from a molten state small organic molecules can transition into amorphous or glassy phases that are commonly known as molecular glasses or amorphous molecular materials[58]. Molecular glasses have broad applications across various industries due to their unique properties. In pharmaceuticals, they enhance solubility and bioavailability of drugs[59-65]. In electronics, they are utilized in technologies like OLEDs[66-72], organic photovoltaics[73-75] and non-linear optics[76-81] offering flexibility and uniformity.

The preparation of molecular glasses primarily involves three techniques: Liquid-quenching methods, spin-coating, and physical vapor deposition (PVD). Liquid-quenching involves rapid cooling which prevents crystallization and results in a metastable glassy state[82-84]. Spin coating is a technique used to deposit uniform layers of organic materials onto substrates by rapidly spinning them, ensuring precise control over film thickness and surface morphology [28, 85-87]. This process is widely employed in the fabrication of organic thin-film devices such as OLEDs[28, 85, 88-92], while PVD entails the evaporation or sublimation of a material in a vacuum chamber, where it condenses onto a substrate to form a glassy thin film[93]. Interest in investigating the properties and applications of vapor-deposited molecular glasses stems from their superior qualities compared to conventional glasses formed through solution processing[28], notably their ultra-stability[22, 29-32, 94]. Vapor-deposited molecular glasses offer higher density[95-97], unique phase transitions[95, 98, 99], improved mechanical properties[100-102] and anisotropy[28, 32, 35, 37, 71, 87, 91, 103-124]. Anisotropy in these glasses is evident in their optical birefringence[28, 37, 106, 108, 113, 117, 118], magnetic behavior[103], and structural characteristics[32, 87, 91, 104, 105, 109, 111, 113, 114].

Understanding the interplay between structural anisotropy and glass stability is critical, as it is influenced by molecular structure and deposition conditions. Structural anisotropy in PVD glasses is largely attributed to preferred molecular orientation[32, 109, 125] and molecular layering[105, 111, 115]. At low deposition temperatures ($T_{dep}$), molecules tend to orient horizontally (parallel to the substrate)[32]. At intermediate $T_{dep}$, elongated molecules typically orient vertically (perpendicular to the substrate)[113]. Deposition near the glass transition temperature ($T_g$), at slow deposition rates, or with molecules possessing smaller aspect ratios, tends to result in isotropic packing[122, 123]. Both experimental and simulation studies indicate that this orientational anisotropy originates from the structure of the supercooled liquid near the surface[35, 37, 113, 123] and stable glasses are formed through surface-mediated equilibrium during PVD[29, 126-131], where molecules near the surface have enhanced mobility allowing them to sample a large number of configurations in search of a more stable state. Thereby, the properties of these glasses can be controlled by adjusting deposition parameters[38, 71, 122, 132].

### Computational challenges

Understanding the relationship between processing conditions and structural properties is essential for optimizing the performance of glassy organic and inorganic films in various technological applications. While computational models based on Molecular Dynamics (MD) and Monte Carlo (MC) simulations offer valuable insights into the properties and behavior of molecular glasses[4, 22-24, 133], there remains a vast gap between these models and real-world laboratory conditions for glass preparation, see Table 1 for the summary of discrepancies in system size and accessible timescales in simulation and experiments.

Most notably, glasses simulations remain plagued by a series of challenges:

i. System size: Due to high computational costs, a significant disparity exists between the system sizes accessible through simulations and those investigated experimentally. Glass simulations are typically limited to relatively small scales, involving systems that range from a few hundred to several thousand molecules/particles[133, 134]. In contrast, experimental studies can probe much larger systems, often extending to macroscopic dimensions encompassing $10^{23}$ molecules/particles or more. This difference highlights a key challenge: simulations, while providing detailed atomic-level insights, are constrained in capturing the full complexity and variability of real glassy materials, which can limit their applicability and accuracy in replicating experimental conditions[134]. Other challenges include insufficient statistical sampling with too few molecules[135, 136], enhanced thermodynamic fluctuations (which scale inversely with the square root of particle count)[137, 138], and errors from limited simulation box sizes that fail to capture large structural features[139, 140]. To balance accuracy and computational feasibility, the system size must be optimized to minimize finite-size effects while maintaining reasonable computational demands.

ii. Limited accessible timescales: MD simulations are constrained to short timescales, typically up to a few microseconds[17], whereas experimental processes can span for hours or days[29, 32, 100]. This disparity results in MD simulations employing ultrafast cooling rates ($10^{14}$–$10^{9}$ K/s), far exceeding those in conventional experiments ($10^{2}$–$10^{0}$ K/s)[17-20]. Consequently, glasses prepared via MD tend to have higher fictive temperatures, making them less stable than experimentally synthesized glasses[20].

For glasses prepared via melt and quench route, the rapid cooling leads to glasses that are less stable and more disordered, remaining in high-energy states. This significant difference in cooling rates between simulations and experiments creates a systematic gap, leading to less stable simulated glasses compared to their experimental counterparts.

On the other hand, for PVD glasses, deposition rates in simulations are also significantly higher (often higher than $10^{8}$ nm/s)[22-27] compared to experimental rates (typically less than $10^{0}$ nm/s)[28-34]. This discrepancy means that simulated films tend to have different microstructures and properties compared to experimentally deposited films[24, 35]. These differences in deposition rates further exacerbate the divergence in film thickness (typically, in simulations, film thicknesses are generally less than 10 nm[22, 25, 26], whereas in experiments, they range from hundreds of nanometers[30, 33, 34] to micrometers[32, 39, 76, 79, 100, 101], see table 1) and structural properties between simulations and experiments, as the rapid deposition does not allow for the same relaxation and ordering processes that occur in experimental conditions[39].

Going beyond mimicking experimental protocols, and adding to the challenge, amorphous structures, with their intricate details influenced by preparation methods, have long posed fundamental questions regarding their most relaxed configuration and associated inherent features. These may or may not correspond to the structural properties of the optimal continuous random network, which is defined as the lowest strain perfectly coordinated large-scale network that can exist. While such structure is, by definition, elusive, it has oriented modelling efforts even when looking at specific materials. For example, early attempts to manually create amorphous silicon transitioned to sophisticated computational approaches, such as the bond-switching method by Wooten, Winer, and Weaire (WWW) for producing as close to perfectly coordinated optimal continuous random network compatible with silicon local bonding environment. Other significant techniques discussed in detail in Part 2 of this review, such as the Activation-Relaxation Technique nouveau (ARTn), Reverse Monte Carlo (RMC), swap Monte Carlo (SMC) have partially advanced our understanding of the relaxation mechanisms and energy landscapes of these materials measuring success both by matching experimental results and closeness to the best available CRNs.

iii. Lack of accurate interatomic interaction potentials: Finally, the accuracy of interatomic interaction models remains limited. The electronic Density Functional Theory (DFT), a class of ab initio methods, is commonly regarded as the gold standard for modeling the potential energy surfaces (PESs) of amorphous materials. However, even at the DFT level, the use of different exchange-correlation functionals can result in significantly different PESs, leading to models with varying structures and properties. Moreover, exploring the DFT PES is computationally expensive, severely limiting the accessible length and time scales. As a result, classical interatomic interaction potentials are often employed, introducing further discrepancies between computational models and real-world amorphous materials.

Empirical forcefields involve optimizing hundreds or thousands of parameters[141, 142]. Even simpler classical forcefields require dozens of parameters, increasing with the number of elements involved[143]. This high dimensionality complicates the identification of what would be the optimal structures based on the values in the cost function landscape of a continuous random network or the best match to experimental results [155]. In addition, the cost function landscape for forcefield optimization is typically rough and filled with numerous local minima[144]. As a result, traditional optimization methods, such as gradient descent, often get trapped in these local minima, making the process highly dependent on the initial starting point[145]. This roughness necessitates numerous independent optimizations and relies heavily on intuition, making the parameterization process biased and laborious[143, 146].

In the last decade, machine-learning (ML) potentials trained on DFT data have helped narrow this gap, see Part 3 in this review for more details. However, challenges remain, particularly concerning interfaces, chemical changes, and charge transfer.

Table 1. Comparative analysis of key aspects between simulations and experiments in the study of molecular glasses

| Aspect | Simulations | Experiments |
| --- | --- | --- |
| System size | $10^2$-$10^6$ molecules | $\sim 10^{23}$ molecules |
| Timescale | $10^{-12}$-$10^{-6}$ s | >10-$10^5$ s |
| Deposition rate | $10^8$-$10^{11}$ nm/s | $10^{-2}$-$10^{-1}$ nm/s |
| Cooling rate | $10^9$-$10^{14}$ K/s | $10^0$-$10^2$ K/s |
| Film thickness | <10 nm | $10^2$-$10^4$ nm |

**Part 2: Beyond mimicking experimental protocols: sampling and the search for optimal structure**

As mentioned before, while the structural details of amorphous configurations depend on the preparation procedures, fundamental questions arose very early on as to the nature and existence of the optimal continuous-random network (CRN) structure, its features, and its generality in representing real systems. Representing thermodynamic ensembles, such structures could be produced through non-dynamical processes driven only by energy or structural deformation (bond length, bond angle and coordination, for example).

If for materials such as silica, a glass characterized by significant topological rigidity near the rigidity transition threshold[146], even short molecular dynamical procedures can lead to low-stress and low-defect structures[147]. Finding low-energy structures in high-connectivity structures is a more significant challenge that amounts to overcoming barriers in a featured free energy landscape, see Figure 2(a) for an illustration. For a pedagogical review on rigidity and mechanical stability of amorphous solids see Ref.[148].

While the reader is referred to the recent review by Laurent J. Lewis[149] for an extensive historical account of modeling amorphous silicon, we focus here on providing a high-level view of the various approaches used to generate high-quality disordered structures beyond the generic quench and melt. In the case of glasses, we similarly refer the reader to the review of Micholaut and Bauchy, which describes how rigidity theory can be employed to study and generate glassy silicate models[150].

Initial attempts to create such structures for amorphous silicon were done 'by hand' by Polk[151] and a few others in early 1970's. This was followed soon after by computer approaches. Among the early success is the ingenious bond-switching approach proposed by Wooten, Winer and Weaire (WWW)[152]to create a disordered network starting from a perfect crystal while preventing the creation of coordination defects using a harmonic Keating potential[153].

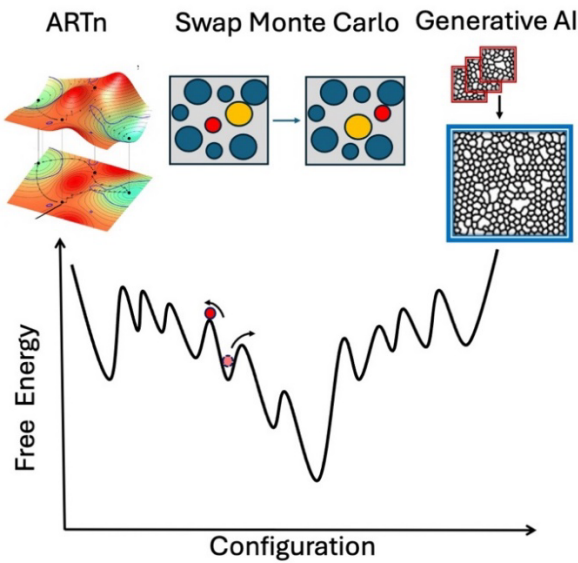 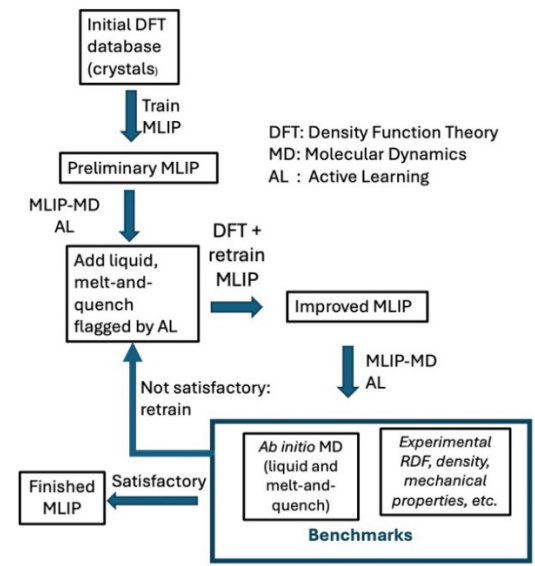

Figure 2. Illustration of computational methodologies used in sampling and potential energy modeling. (a) Featured free energy landscape of amorphous materials requires specialized enhanced sampling techniques such as ARTn, Monte Carlo, and, more recently, approaches based on generative artificial intelligence (AI). (b) Flow-chart of development of machine learning interatomic potentials (MLIPs) using ab-initio methods, molecular dynamics (MD) simulations, and active learning (AL).

The WWW algorithm was used, in its original form[154], and revised versions that greatly accelerated the sampling allowing the generation of high-quality model of 4000[155] to 100,000 atom models that have remained reference models until today[156]. These methods were also used, iterating with relaxations using a modified Stillinger-Weber potential[157] adapted to reproduce amorphous silicon structures[158], to generate near hyperuniform continuous-random network models[159]. While the original WWW algorithm does not impose evenness on the loops connecting atoms, it is possible to modify the method to force even cycles and explore the effects of additional chemical ordering on the structure of binary amorphous semiconductors such as a-GaAs[160], showing the richness of this bond-switching approach to explore fundamental questions about the nature of CRNs. And while the general WWW approach produces the lowest energy structures without coordination defects, the bond-switching moves require crossing high energy barriers that are unphysical. Moreover, experimental evidence even for well-relaxed a-Si shows a significant concentration of low-coordination defects[161] that need to be reproduced by modelling.

In general, defects in amorphous materials include the deviations from a perfect bonding configuration[162], dangling or floating bonds[163], and vacancies[164], local strain-related defects[165] may modify physical and chemical properties of the material[166, 167]. In the past computational studies of defects explored the topological satisfaction of local structures using the atomic-level stresses[164, 168] identified locally favored structures[169], and examined local vibrational modes[170, 171].

Two general approaches were applied to a-Si to address this issue: Reverse Monte Carlo and the Activation-Relaxation Technique. Reverse Monte-Carlo (RMC), first applied to a-Si in 1993[172], aimed at extracting local atomic structure from global experimental averages such as the radial distribution function, using the minimum number of additional constraints. Unlike MD simulations, which depend on interatomic potentials and attempt to replicate real-time dynamics, RMC aims to find atomic configurations that match experimental data without accounting for atomic interactions or thermodynamics [142]. In RMC, atomic positions are randomly adjusted to minimize a cost function representing the difference between the model's pair correlation function and the experimental data. This process follows an accept-reject protocol, similar to the Metropolis Monte Carlo algorithm, but with a key distinction: while Metropolis Monte Carlo uses potential energy as the cost function, RMC uses the discrepancy between the simulated and experimental signals [173, 174] thereby RMC avoids the high cooling rates and approximations associated with MD's forcefields by bypassing the need for melt-quenching. However, its reliance on experimental data limits its predictive power, as it can only replicate structures for conditions that have already been explored [130].

A significant challenge in RMC is that it generates atomic configurations by inverting experimental data, which often results in non-unique solutions. Different atomic arrangements can produce similar or indistinguishable experimental fingerprints, such as $g(r)$[143, 144]. This ambiguity is especially problematic for complex materials like glasses, where multiple valid yet potentially thermodynamically unstable structures may satisfy the same experimental constraints[130, 143, 145]. Through multiple trials, it was discovered that the range of configurations able to reproduce macroscopically averaged experimental data included a large fraction of non-physical configurations and that the inclusion of strict constraints of local configurations was necessary to generate models close to those obtained by quench and melt[175].

Further improving on these methods, Drabold and collaborators introduced the "Force-Enhanced Atomic Refinement" (FEAR) method to improve RMC results by recursively optimizing the structure against local classical forcefield-based[176] or quantum-mechanical DFT-based[177] potentials and global experimental results[178]. While the final configuration meets both constraints, often classical mechanics fails to provide adequate accuracy, and resorting to quantum-mechanical approaches limits the system sizes to only a few hundred atoms[179]. This limits greatly the advantage of this method for generating relevant samples given the local configurational richness of disordered systems such as a-Si. RMC and related techniques are further discussed in Part 3 in the context of using experimental data to inform simulations.

The Activation-Relaxation Technique nouveau (ARTn), an open-ended method for finding local transition states surrounding a local minimum[180, 181], was used to explore the energy landscape, identifying the relaxation mechanisms, and relax the structure of amorphous materials, including a-Si[182, 183], a-GaAs[184] and silica glass[185]. As with molecular dynamics, the resulting low-energy structures correspond to low-energy points of the potential energy used. The validity of the result is therefore determined by the quality of the potential. ARTn, just as the WWW algorithm, does not describe real kinetics, but rather generates activated mechanisms that can be accepted or rejected using, for example, a Metropolis criterion. However, it allows for the generation of a much broader set of mechanisms[186]that provide a better understanding the relaxation mechanisms and provides different pathways to low energy structures, an essential feature to assess the universal properties of optimal CRNs.

In the context of glass transition studies[4] that aim to explore relaxation processes in increasingly viscous substances, Swap Monte Carlo (SMC) pioneered by Grigera and Parisi[187] has become popular as it speeds up relaxation by introducing non-physical moves that involve swapping physically distant particles. Enhanced efficiency is usually attributed to overcoming kinetic barriers typical in glass formers[188], but it also comes with some disadvantages: it has been shown that SMC leads to artificial crystallization in some systems[189, 190]. For instance, Brumer and Reichman found that while SMC was efficient for a two-dimensional hard disk system, 3D polydisperse systems were prone to phase separation/crystallization[189]. Formation of such ordered phases is an unwanted artifact because the resulting structures no longer represent the glassy state of interest. A breakthrough was achieved when Ninarello et al. demonstrated that by carefully tuning the interaction potentials and polydispersity, a set of glass formers can be studied, with speed up thermalization up to 10 orders of magnitudes without crystallization[191]. SMC has since been successfully applied in numerous studies such as measuring the static length scales[192, 193], and generating ultra-stable glasses[194, 195].

Although, as discussed in Ref. [191], SMC's effectiveness is quite restricted, in recent years, adaptive methods augmented by machine learning protocols, such as reinforcement learning and normalizing flows have shown that proposal distributions in Metropolis-Hasting algorithm can enhance the sampling efficiency particularly in systems where the equilibration suffers from slow dynamics and metastability[196, 197]. Thanks to the ongoing intensive method development in the MC community, for example Ref.[198] we expect to see more novel designs of MC methods accompanied by 'smart' moves in context of amorphous materials and glasses in near future.

## Part 3: New ideas: machine learning for improved accuracy, sampling, and reduced mismatch with experiments

*Machine learning interatomic potentials:* Atomistic simulations, powered by machine learning interatomic interaction models,

revolutionize materials science in unprecedented ways[199, 200] lifting many barriers in modeling. None of these barriers, from the spatio-temporal limitation of density function theory to the transferability limitations of semi-empirical models pose a problem for machine learning interatomic potentials (MLIP). While several MLIP models have been developed since the first one by Bheler et al. in 2007[201] they all rely on approximations of the potential energy surface (PES). This approximation assumes a medium without charge or polarization, allowing the total energy of a given system to be approximated by the sum of individual atomic energies. These individual energies strongly depend on the local atomic environment, which is captured by descriptors representing the surrounding atomic configurations[202].

This concept, in principle, is based on atomic environment descriptors, regression methods, and quantum mechanical data[203, 204]. Except for the symmetry of the Hamiltonian, there is no direct parametrization based on the type of physical interactions. Instead, it relies on mathematical interpolation of the potential energy surface (PES), utilizing quantum mechanical data that constitutes a set of discrete points on the PES. Thus, the accuracy of MLIP depends on how well these discrete points cover the targeted PES, a problem that is often addressed by resorting to active learning protocols.

The flow-chart in Figure 2(b) outlines a typical MLIP training protocol. MLIPs differ based on the type of local atomic environment descriptors and regressors used. Current regression methods can be grouped into three main categories: artificial neural networks (NNs)[201, 205], kernel-based method[206], and linear regression[207, 208]. The first two categories have been employed in modeling amorphous silicon (a-Si) and amorphous silica (a-SiO2) using neural network potentials and Gaussian approximation potentials [209-213]. Linear regression potentials include the Spectral Neighbor Analysis Potential (SNAP)[207], atomic cluster expansion[214] and the Moment Tensor Potential (MTP)[208]. These have recently been leveraged in order to create MLIPs able to jointly describe Si, $SiO_2$ and their interfaces[213, 215, 216], in solid, liquid and amorphous states, with a very good level of fidelity.

*Sampling*: Approaches based on machine learning are enriching the search for both experimentally relevant and optimal structures. As mentioned above, progress for the former, was made with the development of DFT-quality machine-learned potentials that made it possible to generate well-relaxed large amorphous models with low defects using quench and melt approaches[206, 217]. Coupled with recent large-scale melt-and-quench work using the modified Stilliger-Weber potential[218], this work demonstrates that with the right potential and sufficient computational effort, it is possible to generate structure that, while presenting some coordination defects, are comparable in structure to the best continuous random networks (CRNs) generated using the Wooten, Winer and Weaire (WWW) approach.

Nevertheless, fundamental questions remain: the minimum strain structure that can be generated - how does it vary with size? After all the work needed to relax a structure to the same level as WWW increases faster than the number of atoms, as seen in Ref. [155], for example; what of binary network, etc.? To answer these questions, Comin and Lewis developed a machine-learning approach to directly generate a-Si structures after learning from high-quality models[219]. If the initial results are still far from optimal, they show that this task is possible with sufficient training data and the right ML model. Indeed, in recent years a variety of generative models have been explored, including GANs refs. [220, 221], autoencoders[219], normalizing flows[222]. Following the work of Comin and Lewis, GAN models have been used to generate amorphous structures based on point cloud representations of molecular input[221].

The fast-decaying structural correlations characteristic of amorphous materials suggest that autoregressive generation may be an effective strategy for sampling large scale amorphous configurations. In an autoregressive approach, the probability of transitions from one microstate to others is inferred from small-scale (order of correlation length) samples of the material. Then the larger sample is extrapolated from small samples, and it is generated one grid-point at a time conditional on previously generated molecular context. The cost of sampling is thereby limited to linear scaling with the number of populated grid points. Recent works in this area developed a modeling approach called the Morphological Autoregressive Protocol (MAP) based on the PixelCNN architecture[223, 224] with grid-based representation of molecular structures showing promising results on systems like amorphous graphene (2D), and liquid water (3D). Grid-based input representation offers the benefits of easy processing and, in the case of 2 dimensional films, the possibility of direct integration of experimental microscopy data into the modeling loop; but comes with high memory demands. As an alternative, point cloud representations offer scalability but require careful design to ensure symmetry invariance[225, 226].

*Informing simulations using experimental data:* The gap between simulation and experiment may be reduced by integrating experimentally measured parameters into simulations. This may be done by including experimental data as simulation parameters [176, 178, 227] or as constraints[228-230]. For instance, molecular dynamics simulations of amorphous silicon (a-Si) and amorphous silica (a-SiO$_2$) are initialized using experimental structural data such as atomic coordinates and lattice parameters[228-231]. These are usually determined experimentally via techniques such as X-ray crystallography[232, 233] and neutron diffraction[234, 235]. Furthermore, a cubic box corresponding to the experimental density of amorphous silica (2.20 g/cm$^3$) was used to prepare the amorphous silica components of the MLIP databases[142, 236, 237].

Experimental data obtained from diffraction, infrared (IR), and nuclear magnetic resonance (NMR) measurements is used constraints to minimize a cost function during the simulation of a-Si and a-SiO2 within the framework of Reverse Monte Carlo (RMC)[238, 239] and related Hybrid Reverse Monte Carlo (HRMC)[240, 241]. For example, experimental structure factors, Si-Si-Si bond angles, and density data were used as constraints to simulate amorphous silicon with the standard Reverse Monte Carlo (RMC) method[242]. In addition, realistic amorphous silica structures may be generated using the RMC method by applying experimental constraints such as Si-O bonds, intra-tetrahedral (O-Si-O) bond angles, and inter-tetrahedral (Si-O-Si) bond angles[243]. The HRMC method encompasses, but is not limited to, Experimentally Constrained Molecular Relaxation (ECMR)[244], Experimentally Constrained Structural Relaxation (ECSR)[245], and Force-Enhanced Atomic Refinement (FEAR)[176]. Both a-Si and a-SiO$_2$ were successfully simulated with high fidelity within these frameworks. For instance, in ECSR, the experimental reduced electron diffraction intensities and the experimental fluctuation electron microscopy FEM variance data were used[245]. The FEAR method was successfully applied to both a-Si and a-SiO2 using the pair distribution function obtained from neutron diffraction and X-ray diffraction data as experimental constraint[176]. These approaches may be applied to molecular glasses, but molecular glasses remain at earlier stages of computational exploration than amorphous silica and silicon[58].

## Part 4: Conclusions and outlook

In this review, we examined the pivotal role of computational simulations in bridging the gap between theoretical predictions and experimental observations. We highlighted significant advancements in modeling amorphous materials, including a-Si, a-SiO$_2$, and molecular glasses, while acknowledging the ongoing challenges related to matching the experimentally relevant system sizes and timescales.

Until recently computational simulations of amorphous materials reflected a choice of either accepting artifacts that arise due to inadequately small size of the simulated systems studied on timescales that are too short using methods that are too expensive to scale up or accepting the artifacts of the inaccuracies that come with reliance on force-fields/purely data-driven techniques. In this review we explored the progress in conformation sampling methods, from techniques that mimic experimental protocols such as the traditional quenching and melting and vapor deposition techniques to more general approaches such as Reverse Monte Carlo (RMC) and the Activation-Relaxation Technique nouveau (ARTn), to the novel machine learning-based approaches developed specifically for amorphous materials, including generative models and autoregressive techniques such as the Morphological Autoregressive Protocol (MAP). Finally, we have discussed a recent development that is of particular importance to the field of amorphous matters – the emergence of machine learning interatomic potentials (MLIPs) that bring DFT-level accuracy to atomistic simulations at a reduced cost relative to DFT, augmenting simultaneously the accuracy and sampling efficiency.

The advances discussed in this review bring us closer to the ultimate goal of informing experiments using computational simulations and guiding and accelerating the design of useful amorphous materials. New ideas, however, are still needed to completely close the gap in time- and length- scales between simulations and experiments. The exciting promise of this field is that by constructing new methods we may unlock an incomprehensively enormous set of new chemical species. And while it may not yet be widely appreciated, to us the future of materials appears to be delightfully amorphous.

## AUTHOR INFORMATION

### Corresponding Author

*Lena Simine – Department of Chemistry, McGill University, Montréal, Québec, H3A 0B8, Canada.

Email: lena.simine@mcgill.ca

### Author Contributions

The manuscript was written through contributions of all authors and all authors have given approval to the final version of the manuscript.

### Funding Sources

National Research Council via the AI4Design program.

## ACKNOWLEDGMENT

We acknowledge funding from the National Research Council via the AI4Design program.